\documentstyle[preprint,tighten,aps]{revtex}
\begin{document}
\draft
\title{\bf Duality symmetry, strong coupling expansion \\
and universal critical amplitudes \\
in two-dimensional $\Phi^{4}$ field models}
\author{ Giancarlo Jug }
\address{
INFM -- UdR Milano, and Dipartimento di Scienze, Universit\`a 
dell'Insubria \\ Via Lucini 3, 22100 Como (Italy) \cite{GJ} \\
and Max-Planck-Institut f\"ur Physik komplexer Systeme,
N\"othnitzer Str. 38 \\ D-01187 Dresden (Germany) }
\author{ Boris N. Shalaev }
\address{ Fachbereich Physik, Universit\"at-Gesamthochschule Essen,\\
D-45117 Essen (Germany) \cite{BNS}}

\maketitle
\begin{abstract}
\noindent

We show that the exact beta-function $\beta(g)$ in the continuous
2D $g\Phi^{4}$ model possesses  the Kramers-Wannier duality symmetry.
The duality symmetry transformation $\tilde{g}=d(g)$ such that 
$\beta(d(g))=d'(g)\beta(g)$ is constructed and the approximate values 
of $g^{*}$ computed from the duality equation $d(g^{*})=g^{*}$ are shown 
to agree with the available numerical results.

The calculation of the beta-function $\beta(g)$ for the 2D scalar
$g\Phi^{4}$ field theory based on the strong coupling expansion is developed 
and the expansion of $\beta(g)$ in powers of $g^{-1}$ is obtained up to 
order $g^{-8}$. 

The numerical values calculated for the renormalized coupling constant
$g_{+}^{*}$ are in reasonable good agreement with the best modern estimates 
recently obtained from the high-temperature series expansion and with those 
known from the perturbative four-loop renormalization-group calculations.

The application of Cardy's theorem for calculating the renormalized 
isothermal coupling constant $g_{c}$ of the 2D Ising model and the related
universal critical amplitudes is also discussed. 

\end{abstract}

\vfill
\newpage

\section{ INTRODUCTION }
\renewcommand{\theequation}{1.\arabic{equation}}
\setcounter{equation}{0}

In this paper we study mainly the symmetry properties of the 
beta-function $\beta(g)$ for the 2D $g\Phi^{4}$ theory, regarded as a 
continuum limit of the exactly solvable 2D Ising model. 
In contrast to the latter, the 2D $g\Phi^{4}$ theory is not an integrable 
quantum field theory. This means, in particular, that the theory does not 
possess the factorized scattering matrix, and therefore that the thermodynamic
Bethe ansatz method cannot be applied at all.

Thus, despite the fact that the 2D Ising model at $h=0$ can be solved by many 
different methods (see \cite{ab1} for an excellent review), the beta-function 
$\beta(g)$ of its continuum limit is to date known only in the four-loop 
approximation within the framework of conventional perturbation theory at 
fixed dimension $d=2$ \cite{ab2,ab3,ab4}. Calculations of beta-functions are 
of great interest in statistical mechanics and quantum field theory. The 
beta-function contains the essential information on the renormalized coupling
constant $g^{*}$, this being important for constructing the equation of state 
of the 2D Ising model -- for example -- which remains still a challenging 
problem, rich in applications. This and other considerations do not allow us 
to regard the 2D Ising model as having fully been solved.

The 2D Ising model and some other lattice spin models are known to possess 
the remarkable Kramers-Wannier(KW) duality symmetry, playing an important 
role both in statistical mechanics and in quantum field theory 
\cite{ab5,ab6,ab7}. The self-duality of the isotropic 2D Ising model means 
that there exists an exact mapping between the high-T and low-T expansions 
of the partition function \cite{ab7}. In the transfer-matrix language this 
implies that the transfer-matrix of the model under discussion is covariant 
under the duality transformation. If we assume that the critical point is 
unique, the KW self-duality would yield the exact Curie temperature of the 
model. This holds for a large set of lattice spin models including systems 
with quenched disorder (for a review see \cite {ab7,ab8}).

Over twenty years ago the KW self-duality was shown to be equivalent to a 
Fourier tranformation in target space \cite{ab9}. Also, it has been recognised
long ago that self-duality combined with some special algebraic properties of 
a model leads to the existence of an infinite set of conserved charges 
\cite{ab10}.  Duality is thus known to impose some important constraints on 
the exact beta-function \cite{ab11,ab12}.

The other main purpose of this paper is to develop a strong coupling expansion
for the calculation of the beta-function of the 2D scalar $g\Phi^{4}$ theory 
as an alternative approach to standard perturbation theory. It will then 
be of interest to match this expansion with the results of a four-loop 
approximation (where possible) by constructing a smooth interpolation with 
respect to $g$. It is in fact well known from quantum field theory and 
statistical mechanics that any strong coupling expansion is closely connected 
with a suitable high-temperature (HT) series expansion for a lattice model 
\cite{ab1,ab7}. From the field-theoretical point of view the HT series are 
nothing but strong coupling expansions for field models, the lattice being 
considered as a technical device to define cutoff-regularised field theories. 

Recently, the high-temperature (HT) series expansions and perturbative
calculations for the $g\Phi^{4}$ field theory at fixed dimensions $d<4$
have been a topic of intense studies (for references see below). Computing 
critical exponents and various critical amplitude ratios from series expansion
data has a long history going back to the early 1960s. Nowadays there are a 
good number of papers containing a large body of information for the 
$N$-vector model defined on different lattices for $d=2,3,4,5$ and arbitrary 
$N$ \cite{ab13,ab14,ab15}. It is remarkable that the HT series data for the 
zero field susceptibility $\chi$ and the second correlation moment $\mu_{2}$ 
of the $N$-component classical Heisenberg ferromagnet have been extended up 
to the order $K^{21} (K=J/T)$, the data for the second field derivative of 
the susceptibility $(\chi_{4})$ being available through to the order $K^{17}$.
Having been equiped with this information, one may try to employ different 
techniques of resummation of the existing HT series expansions, like Pad\'e 
approximants or more subtle approaches, for computing critical exponents and 
universal critical amplitude ratios \cite{ab13,ab16,ab17,ab18,ab19,ab20,ab20*}.
It is worth noting that the strong-coupling behavior of the $g\Phi^{4}$ 
theory has recently been treated within the framework of a variational 
perturbative approach \cite{ab21}. 

The paper is organized as follows. In Sect. II we set up basic notations and 
define the duality symmetry transformation $\tilde{g}=d(g)$. Then it is proved
that $\beta(d(g))=d^{\prime}(g)\beta(g)$. An approximate expression for $d(g)$
providing good estimates for $g^{*}_{+}$ (the renormalised fixed-point 
coupling constant along the isochore line) is found. In Sect. III the HT series
expansion data are used to obtain the strong coupling expansion of $\beta(g)$ 
for the 2D $0(N)$-symmetric $g\Phi^{4}$ theory in powers of $1/g$ up to the 
order $g^{-8}$. Some numerical estimates for the renormalized coupling constant
$g^{*}_{+}$ above $T_{c}$ are obtained. We then compare the fixed point values
found to those already known from the four-loop renormalization-group (RG) 
calculations and from the HT series expansions. In Sect. IV we also discuss 
the application of Cardy's formula both for the exact calculation of the 
renormalized isothermal coupling constant $g^{*}_{c}$ at $T_c$ and, for some 
universal critical amplitudes, along the isothermal critical line. Sect. V 
finally contains some concluding remarks. The Appendix presents a simple 
derivation of the correlation length $\xi$ and of the exact beta-function 
$\beta_{Ising}(T)$ for the lattice 2D Ising model, where the temperature $T$ 
plays the role of an effective coupling constant, and we discuss some of 
their properties.

\section{DUALITY SYMMETRY OF THE BETA-FUNCTION}
\renewcommand{\theequation}{2.\arabic{equation}}
\setcounter{equation}{0}

We begin by considering the classical Hamiltonian of the 2D Ising model
(in the absence of an external magnetic field), defined on a square lattice 
with periodic boundary conditions; as usual:

\begin{equation}
H=-J\sum_{<i,j>}\sigma_{i}\sigma_{j}
\label{b1}
\end{equation}

\noindent
where $<i,j>$ indicates that the summation is over all nearest-neighboring
sites; $\sigma_{i}=\pm1$ are spin variables and $J$ is a spin coupling.
The standard definition of the spin-pair correlation function reads:

\begin{equation}
G(R)=<\sigma_{\bf R}\sigma_{\bf 0}>
\label{b3}
\end{equation}

\noindent
where $<...>$ stands for a thermal average.

The correlation length may be defined in many different ways, all 
definitions being equivalent to each other in the close vicinity of
the critical point \cite{ab13}. This, in fact, reflects the arbitrariness
somewhat inherent in any renormalization scheme.
The statistical mechanics definition of the correlation length is given 
by \cite {ab22}

\begin{equation}
\xi^{2}=\frac{d\ln G(p)}{dp^{2}}|_{p=0}
\label{b4}
\end{equation}

\noindent
The quantity $\xi^2$ is known to be conveniently expressed in terms of the
spherical moments of the spin correlation function itself, namely

\begin{equation}
\mu_{l}=\sum_{\bf R} (R/a)^{l}G({\bf R})
\label{b5}
\end{equation}

\noindent
with $a$ being some lattice spacing. It is easy to see that

\begin{equation}
\xi^{2}=\frac{\mu_{2}}{2d\mu_{0}}
\label{b6}
\end{equation}

\noindent
where $d$ is the spatial dimension (in our case $d=2$). It should be 
mentioned that the above definition of $\xi$ differs from the one used in 
other related approaches, e.g. \cite{ab11}.

The 2D Ising model near $T_{c}$ is known to be equivalent to the
$g\Phi^{4}$ theory with a one-component real order parameter.
In order to extend the KW duality symmetry to the continuous field theory
we have need for a "lattice" model definition of the coupling constant $g$,
equivalent to the conventional one exploited in the RG approach. The 
renormalization coupling constant $g$ of the $g\Phi^{4}$ theory is closely 
related to the fourth derivative of the "Helmholtz free energy", namely
$\partial^{4}F(T,m)/\partial m^{4}$, with respect to the order parameter 
$m=\langle \Phi \rangle$. It may be defined as follows
(see \cite{ab13,ab14,ab23} and references therein)

\begin{equation}
g(T,h)=-\frac{(\partial^{2}\chi/\partial h^{2})}{\chi^{2}\xi^{d}}+
3\frac{(\partial\chi/\partial h)^{2}}{\chi^{3}\xi^{d}}
\label{b7}
\end{equation}

\noindent
where $\chi$ is the homogeneous magnetic susceptibility

\begin{equation}
\chi=\int d^{2}x G(x)
\label{b8}
\end{equation}

\noindent
It is in fact easy to show that $g(T,h)$ in Eq.(\ref{b7}) is merely the
standard four-spin correlation function taken at zero external momenta.
The renormalized coupling constant of the critical theory is defined
by the double limit

\begin{equation}
g^{*}=\lim_{h\rightarrow 0}\lim_{T\rightarrow T_{c}} g(T,h)
\label{b9}
\end{equation}

\noindent
and it is well known that these limits do not commute with each other. 
As a result, $g^{*}$ is a path-dependent quantity in the thermodynamic
$(T,h)$ plane \cite{ab13}.

Here we are mainly concerned with the coupling constant on the isochore 
line $g(T>T_{c},h=0)$ in the disordered phase and with its critical value

\begin{equation}
g^{*}_{+}=\lim_{ T \rightarrow T_{c}^{+}} g(T,h=0)
=-\frac{\partial^{2}\chi /\partial h^{2}}{\chi^{2}\xi^{d}}|_{h=0}
\label{b10}
\end{equation}

\noindent
The couplings $g^{*}_{\pm}$ are of great interest for calculating the
equation of state. Notice that the two different fixed points values 
$g^{*}_{+}$ (defined by the above Eq. (\ref{b10})) and $g^{*}_{-}$ 
(defined by the analogous limit procedure for $T \rightarrow T_{c}^{-}$)
computed above and below the Curie temperature differ vastly from each 
other owing to the broken symmetry of the ordered phase.

The "lattice" coupling constant $g^{*}_{+}$ defined in Eq. (\ref{b10})
is in a given correspondence with the temperature $T_c$. We shall see that 
it will be more convenient to deal with a new variable $s=\exp(2K)\tanh(K)$, 
where $K=J/T$. The standard KW duality tranformation is known to be as 
follows \cite{ab6,ab7}

\begin{equation}
\sinh(2\tilde{K})=\frac{1}{\sinh(2K)}
\label{b11}
\end{equation}

\noindent
It follows from the definition that $s$ transforms as $\tilde{s}=1/s$; 
this implies that the correlation length of the 2D Ising model (see also
the Appendix) $\xi^{2}=\frac{s}{(1-s)^{2}}$ is a self-dual quantity. Now,
on the one hand, we have the formal relation

\begin{equation}
\xi \frac{ds(g)}{d\xi}=\frac{ds(g)}{dg}\beta(g)
\label{b12}
\end{equation}

\noindent
where $s(g)$ is defined as the inverse function of $g(s)$, i.e. $g(s(g))=g$
and the beta-function is given, as usual, by 

\begin{equation}
\xi \frac{dg}{d\xi}=\beta(g)
\label{b13}
\end{equation}

\noindent
On the other hand, it is shown in the Appendix that

\begin{equation}
\xi \frac{ds}{d\xi}=\frac{2s(1-s)}{(1+s)}
\label{b14}
\end{equation}

\noindent
>From Eq.s (\ref{b12}) - (\ref{b14}), a useful representation of the 
beta-function in terms of the $s(g)$ function thus follows

\begin{equation}
\beta(g)=\frac{2s(g)(1-s(g))}{(1+s(g)) \left ( ds(g)/dg \right ) }
\label{b15}
\end{equation}

\noindent
Let us define the dual coupling constant $\tilde{g}$ and the duality
transformation function $d(g)$ as

\begin{eqnarray}
s(\tilde{g})=\frac{1}{s(g)}; \qquad \qquad \tilde{g} \equiv d(g)
=s^{-1}(\frac{1}{s(g)})
\label{b16}
\end{eqnarray}

\noindent
where $s^{-1}(x)$ stands for the inverse function of $x=s(g)$. It is easy to 
check that a further application of the duality map $d(g)$ gives back the 
original coupling constant, i.e. $d(d(g))=g$, as it should be. Notice also 
that the definition of the duality transformation given by Eq. (\ref{b16})
has a form similiar to the standard KW duality equation, Eq. (\ref{b11}).

It is easy to prove that $d^{\prime}(g^{*})=\pm 1$. The maps we are looking  
for have $d^{\prime}(g^{*})=-1$, since the opposite sign leads to the trivial
solution $d(g)\equiv g$. This is also shown in the Appendix.   

Consider now the symmetry properties of $\beta(g)$. We shall see that 
the KW duality symmetry property, Eq. (\ref{b11}), results in the 
beta-function being covariant under the operation $g\rightarrow d(g)$:

\begin{equation}
\beta(d(g))=d^{\prime}(g)\beta(g)
\label{b17}
\end{equation}

\noindent
To prove it let us evaluate $\beta(d(g))$. Then Eq.(\ref{b15}) yields

\begin{equation}
\beta(d(g))=\frac{2s(\tilde{g})(1-s(\tilde{g}))}{(1+s(\tilde{g}))
\left ( ds(\tilde{g}) / d\tilde{g} \right ) } 
\label{b18}
\end{equation}

\noindent
Bearing in mind Eq. (\ref{b16}) one is led to

\begin{equation}
\beta(d(g))=\frac{2s(g)-2}{s(g)(1+s(g))
\left ( ds(\tilde{g}) / d\tilde{g} \right ) }
\label{b19}
\end{equation}

\noindent
The derivative in the r.h.s. of Eq. (\ref{b19}) should be
rewritten in terms of $s(g)$ and $d(g)$. It may be easily done by 
applying Eq. (\ref{b16}):

\begin{eqnarray}
\frac{ds(\tilde{g})}{d\tilde{g}}=
\frac{d}{d\tilde{g}}\frac{1}{s(g)}=-\frac{s^{\prime}(g)}{s^{2}(g)}
\frac{1}{d^{\prime}(g)}
\label{b20}
\end{eqnarray}

\noindent
Substituting the r.h.s. of Eq. (\ref{b20}) into Eq. (\ref{b19}) one 
obtains the desired symmetry relation, Eq. (\ref{b17}).

Therefore, the self-duality of the model allows us to determine the 
fixed point value in another way, namely from the duality equation 
$d(g^{*})=g^{*}$. One may now test the compatibility of this approach 
with the standard methods by computing $d(g)$ in the simplest
approximation. For this purpose let us consider the function $s(g)$ 
given by Eq. (\ref{app10}) (in the Appendix). Making use of a rough 
approximation, one gets   

\begin{eqnarray}
s(g)\simeq\frac{2}{g}+\frac{24}{g^{2}}\simeq\frac{2}{g}
\frac{1}{1-12/g}=\frac{2}{g-12}
\label{b23}
\end{eqnarray}

\noindent
Combining this Pad\'e-approximant with the definition of $d(g)$, 
Eq. (\ref{b16}), one is led to 

\begin{eqnarray}
d(g)=4\frac{3g-35}{g-12}
\label{b24}
\end{eqnarray}

\noindent
The fixed point of this function, $d(g^{*})=g^{*}$, is easily seen to 
be $g^{*}=14$. As expected, $d^{\prime}(g^{*}=14)=-1$.

Before moving to the next topic, we notice that the above-described 
approach may be regarded as another method for evaluating $g^{*}$, fully 
equivalent to the standard beta-function method. A systematic way of 
determining $d(g)$ has not yet been developed, in particular because of 
the specific analytical properties of $d(g)$. Namely, because $g=0$ 
and $g=\infty$ are not regular points of $d(g)$, this function cannot be 
expanded in a Taylor series around these points.

\section{STRONG COUPLING EXPANSION FOR THE BETA-FUNCTION}
\renewcommand{\theequation}{3.\arabic{equation}}
\setcounter{equation}{0}

Our next purpose is to develop a strong coupling expansion for the 2D 
scalar $g\Phi^{4}$ field theory. It has been mentioned already that 
nowadays the HT series expansions for $g(T>T_{c},0)$ are known rather 
well \cite{ab13,ab14,ab15,ab16,ab17,ab18,ab19,ab20}. Eq.s (\ref{app9}) 
and (\ref{app10}), obtained in the Appendix, form the basis of 
our treatment. These equations relate $g$ to $s$, up to the ninth order
in the appropriate expansion parameter. Having been equipped with these 
formulas, one may easily calculate the beta-function $\beta(g)$ as a power 
series in $g^{-1}$. 

Inserting Eq. (\ref{app10}) into Eq. (\ref{b15}) and performing simple 
but somewhat cumbersome calculations, we are led to the desired 
asymptotic expansion for $\beta(g)$  

\begin{eqnarray}
\beta(g) &=&-2g+32-64/g+512/g^{2}+512/g^{3}-30720/g^{4}\nonumber\\
&-&172032/g^{5}+32768/g^{6}-172032/g^{7}+32768/g^{8}+0(g^{-9})
\label{b25}
\end{eqnarray}

\noindent
>From Eq. (\ref{b25}) it follows that in the large-$g$ limit 
$\beta(g)\rightarrow -2g+32$, whilst in the weak coupling regime one 
has for $g\rightarrow 0: \beta(g)\rightarrow +2g$ 
\cite{ab2,ab3,ab4,ab22,ab23}. 
It implies that the continuous function $\beta(g)$ changes 
sign at least once at some fixed point $g^{*}$.

Let us get some numerical estimates for $g^{*}_{+}$ now, from 
Eq. (\ref{b25}), and compare these results with those found from the 
HT series expansions and those of the four-loop RG calculations.
In the standard perturbative approach to quantum field theory at 
fixed dimension one must apply some resummation technique to the
expansions of $\beta(g)$ and other RG-functions. It is interesting
that at least in low orders of perturbation theory the $1/g$-expansion,
Eq. (\ref{b25}), does not require the application of a resummation 
technique. The most reliable numerical estimates of $g^{*}_{+}$ were 
obtained by means of the straightforward solution of the equation 
$\beta(g^{*})=0$, from Eq. (\ref{b25}), taken within the 
$g^{-6}$-approximation (without the last two terms in $g^{-7}$ and 
$g^{-8}$). The five (and best) subsequent approximation are as follows

\begin{eqnarray}
g^{*(1)}_{+}&=&16;\qquad g^{*(2)}_{+}=13.6568;
\qquad g^{*(3)}_{+}=15.0044; \nonumber\\
g^{*(4)}_{+}&=&15.0784; \qquad g^{*(5)}_{+}=14.7632
\label{b26}
\end{eqnarray}

\noindent
Here the index $k$ in $g^{*(k)}_{+}$ indicates that $k+1$ terms are
retained in the fixed point equation under discussion.
These estimates exhibit a regular behavior, the last value being
in very good agreement with the most recent estimate 
$g^{*}_{+}=14.700\pm0.017$ obtained for the square lattice \cite{ab14,ab15}.
The estimates obtained after taking into account the $g^{-7}$ and $g^{-8}$ 
terms differ significantly from the above values.
This is apparently an indication that $1/g$-series also require the 
application of some resummation technique. 

Another approach to obtain a numerical estimate for $g^{*}_{+}$ is
a straightforward solution of Eq. (\ref{app10}), given in the Appendix, 
after setting $s=1$. In contrast to the fixed point equation,  
$\beta(g)=0$, it yields a rather poor value of the renormalized coupling 
constant, $g^{*}_{+}=12.533$, compared to the value reported in 
\cite{ab3,ab14,ab15}.

It is interesting to compare our results with those obtained from the 
beta-function of the 2D Ising model and computed in the four-loop 
approximation, known to provide more or less satisfactory results for 
the critical indices \cite{ab2,ab3,ab4}:

\begin{equation}
\beta(v)= 2v-2v^{2}+1.432346v^{3}-1.861533v^{4}+3.1647764v^{5}+0(v^{6})
\label{b27}
\end{equation}

\noindent
To obtain the beta-function in our normalization we have to change
variables \cite{ab23}

\begin{eqnarray}
g=\frac{8\pi}{3}v;\qquad\qquad \beta(g)=\frac{8\pi}{3}\beta(v)
\label{b28}
\end{eqnarray}

\noindent
The analysis based on the Pad\'e-Borel method of resummation of 
asymptotic series yields $g^{*}_{+}=15.08\pm2.5$ \cite{ab23}, which 
slightly exceeds the best values obtained from the HT series 
calculations: $g^{*}_{+}=14.70\pm0.017$ \cite{ab14,ab15};
$g^{*}_{+}=14.67\pm0.04$ \cite{ab23} (it is tempting to conjecture 
that $g^{*}_{+}=\frac{14\pi}{3}$ \cite{ab23}).

The exact numerical value of the critical exponent 
$\omega=\beta^{'}(g^{*})$ governing the leading corrections to the 
scaling laws \cite{ab22,ab25} is still unknown. The straightforward 
application of the above strong-coupling expansion results, Eq.s (\ref{b25}) 
and (\ref{b26}), yields (within a conventional $\omega >0$ definition)
$\omega=\beta^{'}(g^{*}=14.76)=1.88$. This estimate was obtained without 
exploiting a resummation procedure. It differs vastly from the estimate 
found from the four-loop RG calculations combined with the Pad\'e-Borel 
method: $\omega=1.3\pm0.2$ \cite{ab23}.  This value agrees well with that 
predicted by conformal field theory: $\omega=\frac{4}{3}$ \cite{ab22}.

\section{ISOTHERMAL COUPLING CONSTANT AND CRITICAL AMPLITUDES}
\renewcommand{\theequation}{4.\arabic{equation}}
\setcounter{equation}{0}

The two preceeding Sections were devoted to computing the approximate value 
of the renormalized coupling constant $g_{+}^{*}$ at $h=0$ in the isochore 
limit.  Here we remark that in two dimensions there is a possibility of
calculating the exact value of the renormalized coupling constant $g_{c}^{*}$ 
in the isothermal limit, i.e. at the Curie point in an applied magnetic 
field, namely

\begin{equation}
g^{*}_{c}=\lim_{h\rightarrow 0} g(T=T_{c},h)
\label{b29}
\end{equation}

\noindent
by virtue of Cardy' formula \cite{ab26}. It is in fact essential to stress 
that, in contrast to other isothermal critical amplitudes, $g^{*}_{c}$ is 
fixed by this formula, which reads \cite{ab26}

\begin{equation}
c=3\pi h^{2}(2-\frac{\eta}{2})^{2}\int d^{2}r r^{2}G(r)
\label{b30}
\end{equation}

\noindent
where $c$ is a central charge; $G(r)$ is the two-spin correlation
function; and $\eta$=0.25 is the anomalous scaling dimension of the spin 
variable $\sigma$ in the 2D Ising model. Furthermore, Eq.(\ref{b30})
takes on a more convenient form

\begin{equation}
c=3\pi (4-\eta)^{2} h^{2}\xi^{2}\chi
\label{b31}
\end{equation}

\noindent
and it should be stressed that this formula is valid at $T=T_{c}; h \neq 0$.
At the Curie point the correlation length and the susceptibility are 
described by power laws 

\begin{equation}
\xi(h)=f_{1}^{c}h^{-\frac{2}{4-\eta}};\qquad\qquad
\chi(h)=C_{c}h^{\frac{2\eta-4}{4-\eta}}
\label{b32}
\end{equation}

\noindent
with $f_{1}^{c},C_{c}$ being the isothermal amplitudes. On the one hand, 
from Eq. (\ref{b7}) and Eq. (\ref{b32}) it follows that

\begin{equation}
g_{c}^{*}=\frac{2(2-\eta)(4-3\eta)}{(4-\eta)^{2}C_{c}(f_{1}^{c})^{2}}
\label{b33}
\end{equation}

\noindent
On the other hand, from Eq.s (\ref{b7}) and (\ref{b31}) it is seen  
that the correlation length $\xi$ drops out of the product $g^{*}_{c}c$:

\begin{equation}
cg^{*}_{c}=3\pi(4-\eta)^{2}h^{2} \left ( -\frac{\partial^{2}\chi /
\partial h^{2}}{\chi}+
3\frac{(\partial\chi /\partial h)^{2}}{\chi^{2}} \right )
\label{b34}
\end{equation}

\noindent
Inserting Eq. (\ref{b32}) into the r.h.s. of Eq. (\ref{b34}), one 
obtains the renormalized coupling constant value at the end point
of the isothermal line

\begin{equation}
g^{*}_{c}=\frac{6\pi}{c}(2-\eta)(4-3\eta)
\label{b35}
\end{equation}

\noindent
For the 2D Ising model Eq. (\ref{b35}) yields 
$g^{*}_{c}=273\pi /4$. Eq. (\ref{b31}) is easily seen to impose 
a constraint on the amplitudes $f_{1}^{c}$ and $C_{c}$

\begin{equation}
C_{c}(f_{1}^{c})^{2}=\frac{c}{3\pi (4-\eta)^{2}}
\label{b36}
\end{equation}

\noindent
>From Eq. (\ref{b36}) it follows that what we actually found, by virtue of 
Cardy's formula, is only the product $C_{c}(f_{1}^{c})^{2}$. To compute 
these quantities separately one needs more powerful techniques.

In some seminal papers \cite{ab27,ab28,ab29} it was shown how to compute 
the isothermal amplitudes by making use of the Thermodynamic Bethe 
Ansatz and within the framework  of the form-factor approach. In particular, 
in his paper Fateev obtained the following remarkable result \cite{ab27} 

\begin{equation}
f_{1}^{c}= \frac{\Gamma(2/3)\Gamma(8/15)}{4\sin(\pi/5)\Gamma(1/5)}
[\frac{\Gamma(1/4)\Gamma^{2}(3/16)}{4\pi^{2}\Gamma(3/4)
\Gamma^{2}(13/16)}]^{4/15}=0.2270194675
\label{b37}
\end{equation}

\noindent
>From Eq.s (\ref{b36}) and (\ref{b37}) it follows that
 
\begin{equation}
C_{c}=0.0731998414
\label{b37a}
\end{equation}

\noindent
All these results allow one to compute exactly the two following universal 
combinations \cite{ab13} (see also \cite{ab22} and \cite{ab30})

\begin{eqnarray}
Q_{1}&=&\frac{C^{c}\delta}{(B^{\delta-1}C^{+})^{1/\delta}}\nonumber\\
Q_{2}&=&\frac{C^{+}}{C^{c}}(\frac{f_{1}^{c}}{f_{1}^{+}})^{2-\eta} 
\label{b38}
\end{eqnarray}

\noindent
We recall that the definitions of the critical amplitudes entering 
Eq. (\ref{b38}) are as follows (\cite{ab13,ab22,ab30})

\begin{eqnarray}
M_{s}&=& B (-\tau)^{\beta} \nonumber\\
\chi(T \rightarrow T_{c}+0)&=& C^{+}\tau^{-\gamma} \nonumber\\
\xi &=& f_{1}^{+}\tau^{-\nu}
\label{b39}
\end{eqnarray}

\noindent
$M_{s}$ being the spontaneous magnetization. The amplitudes 
$f_{1}^{c},C_{c}$ have been defined in Eq.(\ref{b32}). The exact values 
are as follows: $\delta=15, \eta=0.25, \nu=1, \beta=0.125, \gamma=1.75$, 
$B=1.222410$ (\cite{ab13}), $f_{1}^{+}=0.567296$ (\cite{ab13}, see also 
the Appendix), and  $C^{+}=0.962582$ (\cite{ab31}). Substituting all 
these values into Eq. (\ref{b38}) yields 
  
\begin{eqnarray}
Q_{1}&=& 0.912648 \nonumber\\
Q_{2}&=& 2.647714 
\label{b40}
\end{eqnarray}

The exact values found provide a good opportunity to test the numerical 
results obtained from the HT series expansions and from Monte Carlo 
simulations. The fair estimates obtained from the analysis of HT series 
in the 2D Ising model on the square lattice yield  $f_{1}^{c}=0.233$
and $C_{c}=0.0706$ (\cite{ab13}), whilst the exact results are given by  
Eq.s (\ref{b37}) and (\ref{b37a}). As for the universal combinations 
$Q_{1,2}$, the series expansion analysis yields $Q_{1}=0.88023, Q_{2}=2.88$ 
\cite{ab13}.

Notice in conclusion that Eq.s (\ref{b35}) and (\ref{b36}) hold good 
also for the general case of the 2D $g\Phi^{4}$ $0(N)$-symmetric model
for $-2<N<2$, in particular for the minimal models of conformal 
field theory corresponding to the discrete values of $N$:   
$N=2\cos(\frac{\pi}{m}); m=3,4,...,\infty$. More interesting still is 
that certain combinations of isothermal critical amplitudes, in particular 
$C_{c}(f_{1}^{c})^{2}$, have the same numerical values both in the pure and 
in the quenched disordered 2D Ising model where $c=0.5; \eta=0.25$  
\cite{ab8}.

\section{CONCLUDING REMARKS}
\renewcommand{\theequation}{5.\arabic{equation}}
\setcounter{equation}{0}

We have proved the existence of the duality symmetry transformation $d(g)$
in the $2D$  $g\Phi^{4}$ theory such that $\beta(d(g))=d^{\prime}(g)\beta(g)$.
Actually, this symmetry property was shown to result from the KW duality 
of the 2D lattice Ising model.

It would be tempting but wrong to regard $d(g)$ as a function connecting 
the weak-coupling and strong coupling regimes. For instance, this phenomenon 
takes place in the $\sinh$-Gordon theory with $\beta(d(g))\equiv 0$, the 
exact $S$-matrix being invariant under the strong-weak coupling duality 
$g\rightarrow 8\pi /g$ \cite{ab28}.
 
As a matter of fact, our proof is based on the properties of $g(s),s(g)$ 
defined only for $ 0\leq s < \infty; g^{*}\leq g< \infty$ and therefore 
doesn't cover the weak-coupling region, $0\leq g \leq g^{*}$. Whether the 
beta-function $\beta(g)$ does have the dual symmetry in the weak-coupling 
region, remains an open question.

In contrast to widely held views, the duality symmetry imposes only mild 
restrictions on $\beta(g)$. It means that this symmetry property fixes 
only even derivatives of the beta-function $\beta^{(2k)}(g^{*}) (k=0,1,...)$ 
at the fixed point, leaving the odd derivatives free. In fact, the 
duality equation $d(g)=g $ provides yet another method for determining the 
fixed point, independently of the approach based on the equation $\beta(g)=0$.
Another open problem is also that of finding a systematic approach for 
calculating $d(g)$. 

In this paper the strong coupling expansion for the beta-function of the 2D
Ising model has been developed. This approach was shown to provide reliable 
results for the numerical value of $g^{*}_{+}$ which are found to be in 
good agreement with the best results from the HT series expansion. It is 
worth noting that $g^{*}_{+}$ remains the same for the 2D Ising model with 
random bonds \cite{ab8}.

The method presented in this paper may be easily extended to the 2D 
$0(N)$-symmetric $g\Phi^{4}$ theories for arbitrary $N$. However, the above 
approach appears to be unable to provide reliable numerical results for 
the critical exponent $\omega$, at least without the combined use of a 
resummation technique. In fact, the results obtained disagree both with 
estimates found from the four-loop RG calculations, combined with the 
Pad\'e-Borel resummation method \cite{ab23}, and with the exact value 
$\omega=4/3$ given by conformal field theory \cite{ab22}. As a matter of 
fact, the situation with the critical exponent $\omega$ looks somewhat 
unresolved. On the one hand, all the corrections to the scaling laws 
in the 2D Ising model are analytical. For instance, the susceptibility 
near $T_{c}$ is given by $\chi=C^{+}\tau^{-7/4}+C^{+}_{1}\tau^{-3/4}+...$ 
\cite{ab31}. On the other hand, corrections to scaling are known to be 
powers of $\tau^{\omega\nu}$ \cite{ab22}. All this would lead to 
$\omega=1$, in obvious contradiction to conformal field theory.
Moreover, the spectrum of conformal dimensions of the 2D Ising model 
consists of just three numbers, these being  $(0, \frac{1}{8}, 1)$. 
That is the reason why the appearence of an operator with a fractional 
scaling dimension $4/3$ is so far unclear \cite{ab22}. Thus, although 
one can make a meaningful comparison between the numerical values obtained, 
nevertheless the apparent spread of these numerical results for 
$\omega$ does not allow one to construct a smooth interpolation
for $\beta(g)$ between the two regions: $g<g^{*}_{+}$ and $g>g^{*}_{+}$.

Regarding critical amplitudes, namely, $f_{1}^{c},C_{c},Q_{1}, Q_{2}$, 
we can conclude that the known numerical results are in a surprisingly 
good agreement with the exact results. For further discussion, see 
Ref. \cite{ab33}.  

We have already seen that the exact beta-function $\beta_{Ising}(T)$ 
is essentially non-perturbative. Non-perturbative terms give rise to 
nontrivial fixed points in contrast to the standard beta-functions of 
non-linear sigma models with continuous symmetry, in particular for 
the $0(N)$-symmetric theory with $N>2$  \cite{ab32}. The very existence 
of non-perturbative terms is indicative of the mathematical illegitimacy 
of the naive analytical continuation from the $N>2$ to the $N<2$ region 
often employed in statistical mechanics and condensed matter physics.

\section{ACKNOWLEDGEMENTS}

This work was supported by the Russian Foundation for Basic Research
Grant No. 98-02-18299, the NATO Collaborative Research Grant No.
OUTR.CRG960838 and by the EC contract No. ERB4001GT957255. One of 
the authors (GJ) is most grateful to the Max-Planck-Institute f\"ur 
Physik komplexer Systeme, Dresden, where part of this work was carried 
out, for kind hospitality and the use of its facilities. The other 
author (BNS) is most grateful to the Department of High Energy Physics
of the International School for Advanced Studies in Trieste and, 
especially, to Fachbereich Physik Universit\"at GH Essen, where this 
work was completed, for support and exceptionally warm hospitality.
He has much benefitted from numerous helpful discussions with H.W.Diehl, 
A.I.Sokolov, S.N.Dorogovtsev, Y.V.Fyodorov, Yu.M.Pis'mak, and K.J.Wise. 
He is deeply grateful to G.Mussardo, P.Pujol and A.Honecker for 
interesting discussions on Cardy's formula.

\section{APPENDIX}
\renewcommand{\theequation}{A.\arabic{equation}}
\setcounter{equation}{0}

Here we begin with giving a simple derivation of the correlation length 
$\xi$ for the 2D Ising model from the high-temperature expansions and will 
then present some useful relations.
 
The HT expansions up to $K^{8} (K\equiv \frac{J}{T})$ are as follows 
\cite{ab15}

\begin{eqnarray}
\chi&=&1+4K+12K^2+104K^{3}/3+92K^{4}+3608K^{5}/15\nonumber\\
&+&3056K^{6}/5+484528K^{7}/315+400012K^{8}/105
\label{app1}
\end{eqnarray}

\begin{eqnarray}
\mu_{2}&=&4K+32K^{2}+488K^{3}/3+2048K^{4}/3+38168K^{5}/15\nonumber\\
&+&394624K^{6}/45+8994736K^{7}/315+28064768K^{8}/315
\label{app2}
\end{eqnarray}

\begin{eqnarray}
\chi^{\prime \prime}_{hh}&=&-2-32K-264K^{2}-4864K^{3}/3-8232K^4
-553024K^{5}/15\nonumber\\
&-&2259616K^{6}/15-180969728K^{7}/315-217858792K^{8}/105
\label{app3}
\end{eqnarray}

\noindent
with $\chi^{\prime \prime}_{hh}$ being the second derivative of the 
homogenous susceptibility with respect to a magnetic field $h$; $\chi$ 
and $\mu_{2}$ were defined in Sect. II.

The standard RG equation for the effective temperature $T$ is given by

\begin{equation}
\xi \frac{dT}{d\xi}=\beta_{Ising}(T)
\label{app4}
\end{equation}

\noindent
where $\xi$ is the correlation length (Eq. (\ref{b4}))

\begin{equation}
\xi^{2}=\frac{\mu_{2}}{2d\chi}
\label{app5}
\end{equation}

\noindent
The key observation for computing  $\xi$ is to make use of the new variable 
$s=\exp(2K)\tanh(K)$. One has to substitute Eq.s (\ref{app1}) and 
(\ref{app2}) into Eq. (\ref{app5}) and then to rewrite the expression 
obtained in terms of $s$.  At this order of approximation the procedure 
gives the result 

\begin{eqnarray}
\xi^{2}&=&s+2s^{2}+3s^{3}+4s^{4}+5s^{5}+6s^{6}+7s^{7}+8s^{8}+0(s^{9})
\nonumber\\
&=&\frac{s}{(1-s)^{2}}
\label{app6}
\end{eqnarray}

\noindent
We believe the closed-form above to be an exact result (it satisfies in
particular the $s\rightarrow s^{-1}$ duality symmetry). However, it
should not be confused with other exact formul{\ae} available in the 
literature for $\xi$, based for instance on the transfer-matrix eigenvalue 
approach \cite{ba00} rather than our field-theoretic definition.

One sees that the correlation length as given by Eq. (\ref{app6}) in its
closed form does exhibit the correct behavior near the Curie point

\begin{eqnarray}
\xi&=&f_{1}^{+}\frac{1}{\tau};\qquad \tau=\frac{T-T_{c}}{T_{c}};
\qquad T_{c}=\frac{2}{\ln(\sqrt{2}+1)}=2.269185 \nonumber\\
f_{1}^{+}&=& \frac{1}{4\ln(\sqrt{2}+1)}=0.567296
\label{app7}
\end{eqnarray}

\noindent
where $f_{1}^{+}$ is the correlation length critical amplitude \cite{ab13}.

Combining Eq.s (\ref{app4}) and (\ref{app6}) and carrying out some
calculations, we arrive at a nice expression for the beta-function
written in terms of $s$

\begin{eqnarray}
\frac{ds}{d\ln\xi}=\beta_{Ising}(s)=\frac{2s(1-s)}{1+s}
\label{app8}
\end{eqnarray}

\noindent
Substituting now Eq.s (\ref{app1}), (\ref{app2}) and (\ref{app3}) 
into the r.h.s. of Eq. (\ref{b10}) and making use of the variable $s$, 
one is led to some helpful relations linking $s$ and $g$

\begin{eqnarray}
\frac{1}{g}&=&s/2-3s^{2}+39s^{3}/2-138s^{4}+2061s^{5}/2\nonumber\\
&-&7909s^{6}+122907s^{7}/2-480012 s^{8}/945+0(s^{-9})
\label{app9}
\end{eqnarray}

\begin{eqnarray}
s&=&2/g+24/g^{2}+264/g^{3}+2976/g^{4}+35136/g^{5}+423680/g^{6}\nonumber\\
&+&5149824/g^{7}+63275520/g^{8}+0(g^{-9})
\label{app10}
\end{eqnarray}

\noindent
Rewritten in terms of $T$, Eq. (\ref{app8}) reads

\begin{eqnarray}
\frac{dT}{d\ln\xi}=\beta_{Ising}(T)=
T^{2}\frac{1-2\exp(-2/T)-2\exp(-4/T)+2\exp(-6/T)+\exp(-8/T)}
{1+2\exp(-2/T)+2\exp(-6/T)-\exp(-8/T)}
\label{app11}
\end{eqnarray}

\noindent
The exact beta-function given by Eq.s (\ref{app8}) and (\ref{app11}) 
exhibits the following properties:

(i) it vanishes just at two fixed points (FP): at the Gaussian FP $s=0$
and at the Ising FP $s=1$;

(ii) the derivative of the beta-function at the Ising FP gives the 
critical exponent of the correlation length $\xi$, more precisely

\begin{eqnarray}
\nu^{-1}=-\frac{d\beta_{Ising}(s)}{ds}_{|s=1}=1
\label{app12}
\end{eqnarray}

(iii) the beta-function is covariant under the duality transformation, 
namely

\begin{equation}
\beta_{Ising}(s)=-s^{2}\beta_{Ising}(\frac{1}{s})
\label{app13}
\end{equation}

\noindent
and this equation has been referred to as the consistency condition for 
the beta-function \cite{ab11};

(iv) these properties fully agree with results from the HT series 
expansions.

\noindent
The duality property (iii) implies that the RG equation under
discussion takes the same form both in the dual and in the original 
variables. We note in passing that the well-known self-dual beta-function 
obtained within the Migdal approximation for $d=2$, which reads \cite{ab24}
 
\begin{equation}
\beta_{Migdal}(T)=-T-\frac{T^{2}}{2}\sinh\frac{2}{T}\ln\tanh\frac{1}{T}
\label{app14}
\end{equation}

\noindent
does not satisfy either properties (ii) and (iv) \cite{ab24}.

To end with, we prove some useful relations concerning the duality 
transformation $\tilde{g}=d(g)$ introduced in Sect. II.

i) Let us show that $d^{\prime}(g^{*})=\pm 1$. First of all, from the
definition

\begin{equation}
d(g)\equiv s^{-1}(\frac{1}{s(g)})
\label{e1}
\end{equation}

\noindent
we have that $d(d(g))=g$. Differentiating this with respect to $g$ 
one obtains 

\begin{eqnarray}
d(d(g))&=&g; \qquad \qquad d(g^{*})=g^{*}\nonumber\\
d^{\prime}(d(g))d{\prime}(g)&=&1;\qquad  \qquad  g=g^{*}\nonumber\\
d{\prime}(g^{*}) d{\prime}(g^{*})&=&1
\label{e2}
\end{eqnarray}

\noindent
showing that indeed $d{\prime}(g^{*})=\pm 1$.

iii) Differentiating Eq.(\ref{e2}) (second from top) with respect to 
$g$ one obtains 

\begin{eqnarray}
d^{\prime}[d(g)]d{\prime}(g)&=&1;\nonumber\\
d^{\prime\prime}[d(g)]d^{\prime}(g)^{2}
+d^{\prime}[d(g)]d^{\prime\prime}(g)&=&0
\label{e3}
\end{eqnarray}

\noindent
and at the fixed point we arrive at ($d(g^{*})=g^{*}$):

\begin{equation}
d^{\prime \prime}(g^{*})(d(g^{*}))^{2}
+d^{\prime}(g^{*})d^{\prime\prime}(g^{*})=0
\label{e4}
\end{equation}

\noindent
If $d^{\prime}(g^{*})=-1$ we have an identity 
$d^{\prime \prime}(g^{*})=d^{\prime \prime}(g^{*})$. In the opposite case 
from $d^{\prime}(g^{*})=+1$ it follows that $d^{\prime \prime}(g^{*})=0$. 
Proceeding in the same way, it is easy to see that all higher derivatives 
vanish identicatically at the fixed point. This, alone, does not imply that 
$d(g)\equiv g$, since it could be that $d(g)=g+f(g)$ where $f(g)$ is some 
nonanalytic function having vanishing derivatives at $g^{*}$. Let us then 
assume that $f(g)$ is the first term of an asymptotic expansion of $d(g)$ 
around the fixed point, so that $f(g)\rightarrow 0$ as $g\rightarrow g^{*}$. 
Then remembering that $d(d(g))=g$, 

\begin{equation}
g=d(d(g))=d(g+f(g))=g+f(g)+f(g+f(g))\simeq g+ 2f(g)
\label{e5}
\end{equation}

\noindent
and we do obtain $f(g) \equiv 0$.

\end{document}